\documentclass[preprint,showpacs,preprintnumbers,amsmath,amssymb]{revtex4}

\usepackage{graphicx}
\usepackage{dcolumn}
\usepackage{bm}
\usepackage{amsmath}

\begin{document}


\title{Configuration mixing in $^{188}$Pb: band structure and electromagnetic properties}

\author{V. Hellemans}
\author{R. Fossion}
\author{S. De Baerdemacker}
\author{K. Heyde}

\affiliation{Department of Subatomic and Radiation Physics, Proeftuinstraat 86, B-9000 Gent, Belgium}

\begin{abstract}
\noindent In the present paper, we carry out a detailed analysis of the presence and mixing of various families of collective bands in $^{188}$Pb. Making use of the interacting boson model, we construct a particular intermediate basis that can be associated with the unperturbed bands used in more phenomenological studies. We use the E2 decay to construct a set of collective bands and discuss in detail the B(E2)-values. We also perform an analysis of these theoretical results (Q, B(E2)) to deduce an intrinsic quadrupole moment and the associated quadrupole deformation parameter, using an axially deformed rotor model.
\end{abstract}

\maketitle
\section{Introduction}
\noindent The lead isotopes provide a unique laboratory to study the phenomenon of shape coexistence in nuclei \cite{KHEY1983,JWOO1992, RJUL2001,AAND2000} and have very recently been subject of much experimental and theoretical interest \cite{AFRA2004,MBEN2004,RROD2004,JEGI2004,GDRA2004,MBEN2003,AWAL2003,KVEL2003,GDRA2003,WREV2003,NSMI2003,TDUG2003}. The combined effect of the proton shell-gap at $Z$=82 for the Pb-nuclei and the large number of valence nucleons outside the closed $N$=126 core (in this case, neutron holes), results in an important lowering of the energy of proton particle-hole excitations \cite{KHEY1987}. More specifically, near neutron midshell (at $N$=104), the proton 2p-2h and 4p-4h excitations descend to very low  excitation energy because of the very large proton-neutron binding energy that results from the interactions between the core-excited protons across the $Z$=82 shell closure and the large number of valence neutrons outside of the $N$=126 shell closure. As a consequence, mixing can result between various families of configurations having approximately the same excitation energy. More in particular, the lowest-lying $0^+$ and $2^+$ states can become strongly mixed such that it is very difficult to assign a 'configuration label' to them. In section \ref{sect:mix}, we succintly describe the essentials of the configuration mixing using the interacting boson model (IBM) \cite{FIAC1987,AFRA1994,PDUV1982}.In section \ref{sect:be2values} we have in mind to obtain a better understanding in the nature of the low-spin collective states and to construct particular bands by making use of the calculated E2 decay probabilities starting at the high-spin states in the particular case of $^{188}$Pb. In a final section \ref{sect:rotor}, we use the calculated values (using the IBM) for quadrupole moments and B(E2)-values in these bands in order to extract collective model parameters (intrinsic quadrupole moments, quadrupole deformation $\beta_0$) in order to highlight the equivalence between the IBM approach, used as a highly-truncated shell-model calculation, and a geometrical rotational model, as can be derived from mean-field methods \cite{MBEN2004,JEGI2004,TDUG2003,MBEN2003,RBEN1989,WNAZ1993}.  
\section{Configuration mixing}\label{sect:mix}
\noindent In a recent study that concentrated on describing intruder bands and configuration mixing in the neutron
deficient Pb-isotopes, a three-configuration mixing calculation has been performed in the context of the  interacting boson model (IBM) \cite{RFOS2003}. We refer to that paper for more details but present succintly the main points. Approximating the regular and intruder states as built from proton 0p-0h, 2p-2h and 4p-4h excitations across the $Z$=82 proton closed shell mainly that interact with the large number of valence neutrons outside of the N=126 neutron closed shell, the Hamiltonian within the IBM takes the form
\begin {equation}
\hat{H}=\hat{H}_{reg}+\hat{H}_{2p-2h}+\hat{H}_{4p-4h}+\hat{V}_{mix}~,\label{eq:hamiltonian}
\end{equation}
with
\begin{eqnarray}
\hat{H}_{reg}&=&\varepsilon_{reg}\hat{n}_d +\kappa_{reg}\hat{Q}_{reg}\cdot\hat{Q}_{reg}~,\\
\hat{H}_{i}&=&\varepsilon_{i}\hat{n}_d+\kappa_{i}\hat{Q}_{i}\cdot\hat{Q}_{i}+\Delta_i~,
\end{eqnarray}
and 
\begin{eqnarray}
\hat{V}_{mix}&=&\hat{V}_{mix,1}+\hat{V}_{mix,2}~,\\
\hat{V}_{mix,i}&=&\alpha_i \left(s^{\dag}\cdot s^{\dag}+s\cdot s\right)+\beta_i \left( d^{\dag}\cdot d^{\dag}+\tilde{d}\cdot\tilde{d}\right)~.
\end{eqnarray}
The quadrupole operator $\hat{Q}_i$ is defined as
\begin{equation}
\hat{Q}_i=\left(s^\dag\tilde{d}+d^\dag\tilde{s}\right)^{(2)}+\chi_i\left(d^\dag\tilde{d}\right)^{(2)}~.
\end{equation}
For details on the notation we refer to Fossion et al. \cite{RFOS2003}. The diagonalisation of the energy matrices, corresponding to the Hamiltonian \eqref{eq:hamiltonian}, is carried out in the $U(5)$-basis, expressing the eigenvectors in the [$N$]$\oplus$[$N+2$]$\oplus$[$N+4$] model space. Unfortunately, using this method, one has no clear insight in the amount of mixing between the unperturbed bands that result from a diagonalisation in the separate subspaces [$N$], [$N+2$], and [$N+4$], respectively. \\\\
\noindent So instead of a complete diagonalisation of the Hamiltonian matrix, we first rotate into an intermediate basis in which the separate parts of the Hamiltonian \eqref{eq:hamiltonian}, i.e. $\hat{H}_{reg}$, $\hat{H}_{2p-2h}$, and $\hat{H}_{4p-4h}$ only, become diagonal. 
We thus have three different bases: \\
\begin{itemize}
\item The \textit{U(5)-basis} $|J,k\rangle_N$, with $J$ the angular momentum, $N$ the number of bosons and $k$ the rank number,
\item the basis in which the full Hamiltonian \eqref{eq:hamiltonian} is diagonal $|J,i\rangle$, and, 
\item the \textit{intermediate bases} in which the Hamiltonian \eqref{eq:hamiltonian} (excluding $\hat{V}_{mix}$) is diagonal in the three different subspaces [$N$],[$N+2$], and [$N+4$]. Respectively they are denoted as $|J,l\rangle'_{N}$, $|J,l\rangle'_{N+2}$, and $|J,l\rangle'_{N+4}$, with $l$ also a rank number.
\end{itemize}
The above bases are connected in the following way. The basis in which the full Hamiltonian is diagonal (expressed in the $U(5)$-basis) reads
\begin{align}
|J,i\rangle &= \sum_{k=1}^{dim_N} a_{k,i}^{N}(J)|J,k\rangle_N\nonumber\\
&+\sum_{l=dim_N+1}^{dim_N+dim_{N+2}} a_{l,i}^{N+2}(J)|J,l\rangle_{N+2}\nonumber\\
&+\sum_{m=dim_N+dim_{N+2}+1}^{dim_N+dim_{N+2}+dim_{N+4}} a_{m,i}^{N+4}(J)|J,m\rangle_{N+4}~,\label{eq:full}
\end{align}
and the `intermediate' basis becomes
\begin{align}
&|J,l\rangle'_N=\sum_{k=1}^{dim_N}b_{k,l}^{N}(J)|J,k\rangle_N~,\label{eq:intermediate}
\end{align}
(similarly for $N+2$ and $N+4$). Dim$_N$, dim$_{N+2}$, and dim$_{N+4}$ are the dimensions of the corresponding configuration spaces for a certain angular momentum $J$ containing $N$, $N+2$, and $N+4$ bosons respectively. The matrix $\mathbf{B^{N}}$ diagonalises the configuration with $N$ bosons and the matrix $\mathbf{A}$ diagonalises the full Hamiltonian \eqref{eq:hamiltonian} directly (the indices $N$, $N+2$ ,and $N+4$ for matrix $\mathbf{A}$ are only added for the sake of clearness). We omit the dimensions of the summations from now on. 

\begin{figure}[!ht]
\begin{center}
\includegraphics[height=8.6cm,angle=-90]{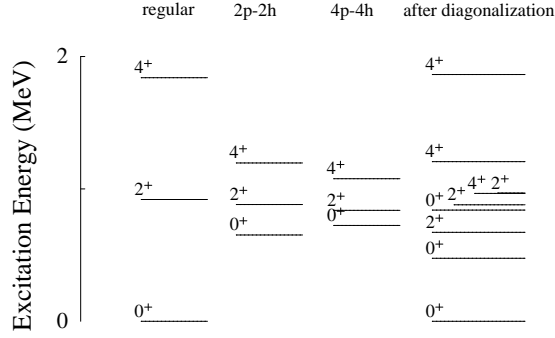}
\caption{Low-energy part of the IBM-spectrum for $^{188}$Pb. The three configurations taken from the left show the absolute energies for the lowest unperturbed bands. At the extreme right, the IBM spectrum after diagonalisation of the full Hamiltonian (\ref{eq:hamiltonian}) is shown. }\label{fig:ibmspectrum}
\end{center}
\end{figure}

\noindent Rotation of the Hamiltonian matrix expressed in the $U(5)$-basis into the `intermediate' basis results in the energy levels of a set of bands in the 0p-0h, 2p-2h and 4p-4h subspaces separately (see also left part in Fig. \ref{fig:ibmspectrum}). These bands correspond to the unperturbed bands that are extracted in phenomenological calculations as carried out by Dracoulis \textit{et al.}\cite{GDRA2004}, Allatt \textit{et al.} \cite{RALL1998}, and Page \textit{et al.} \cite{RPAG2003}. From now on we will call the energy levels (bands) resulting from rotation of the Hamiltonian matrix into the `intermediate' basis unperturbed levels (bands) in order to avoid confusion and we will denote them as the $|J,L\rangle'_N$ states (see eq. \eqref{eq:intermediate}). \\
\noindent By calculating the mixing matrix elements of $\hat{V}_{mix}$ in this `intermediate' basis, we obtain the mixing matrix elements for all unperturbed levels. The knowledge of these unperturbed bands and their mixing matrix elements makes the process of configuration mixing in a nucleus more transparent than before. Starting from the experimental level energies at high spin and the corresponding moment of inertia, a set of unperturbed experimental bands can be deduced extrapolating to the low-spin members of these bands \cite{GDRA2004,GDRA1994,GDRA1998,RALL1998,RPAG2003}. The comparison between the IBM unperturbed bands and these extrapolated unperturbed experimental bands, together with the knowledge of the full energy spectrum and the B(E2)-values form an extensive test for the parameters that describe a certain isotope chain. \\\\
\noindent Starting from IBM-parameters for the Pb-isotopes as determined by Fossion \textit{et al.} \cite{RFOS2003}, a slightly different fit was performed. The parameters for $\hat{H}_{reg}$, $\hat{H}_{2p-2h}$, and  $\hat{H}_{4p-4h}$ remain unchanged, except for $\varepsilon_{reg}$ which was taken 0.92 MeV instead of 0.90 MeV. The mixing parameters $\alpha_i$ and $ \beta_i$ were fixed in $^{196}$Pb. The value for $\Delta_1$ was obtained as the result of a fitting procedure for the Pb-isotopes ($A$=186-196). Basically $\Delta_1$ was fitted for $^{186}$Pb and $^{196}$Pb and the $\Delta_1$ for the other isotopes was chosen following a linear variation between $\Delta_1$($^{186}$Pb) and $\Delta_1$($^{196}$Pb). Then, the difference between the experimental $0^+_2$ and the IBM $0^+_2$ was taken for all isotopes considered and added to the corresponding $\Delta_1$. This method gives a better description of the slope of the energy levels through the isotope chain. Moreover, the $\Delta_1$($^{188}$Pb) obtained in this way is in good agreement with the theoretical prediction that makes use of experimental separation energies \cite{KHEY1987}. Since $\Delta_2$ is associated with the unperturbed energy to excite 4p-4h configurations, the value was taken as 2$\cdot \Delta_1$ \cite{KHEY1991}. The parameters for $^{188}$Pb are listed in Table~\ref{tab:parameters}.
\begin{table}
\begin{center}
\begin{tabular}{lll}
\hline\hline $\alpha_{1}=\beta_{1}$ & $\alpha_{2}=\beta_{2}$ & $\Delta_{1}$\\
\hline 8.5 keV & 23.4 keV & 1923 keV\\
\hline\hline
\end{tabular}
\caption{IBM-parameters used for $^{188}$Pb.} \label{tab:parameters}
\end{center}
\end{table}
The low-energy part of the resulting IBM-spectrum for $^{188}$Pb is presented in Figure~\ref{fig:ibmspectrum}. The comparison with the experiment will be discussed in the next section.\\\\
\noindent By inspecting the unperturbed lowest bands (see left part of Fig. \ref{fig:ibmspectrum}) one notices that mixing modifies the nature of the 2p-2h and 4p-4h $0^+$ states and the three $2^+$ states because of the small energy differences between these unperturbed states (see Figure \ref{fig:ibmspectrum}). The interaction matrix elements between the first three $0^+$ states and the first three $2^+$ states are given in Table \ref{tab:mmelements}.
\begin{table}
\begin{center}
\begin{tabular}{c|c|c|c}
~& $0_1^+$ & $0_2^+$ & $0_3^+$\\
\hline $0_1^+$ & 0 & -0.0768 & 0\\
\hline $0_2^+$ & -0.0768 & 0 & 0.1908\\
\hline $0_3^+$ & 0 & 0.1908 & 0\\

\end{tabular}
\begin{tabular}{c c c }
~&~&~ \\
~ & ~&~ \\
~ & ~&~ \\
~& ~&~ \\
\end{tabular}
\begin{tabular}{c|c|c|c}
~& $2_1^+$ & $2_2^+$ & $2_3^+$\\
\hline $2_1^+$ & 0 & 0.0711 & 0\\
\hline $2_2^+$ & 0.0711 & 0 & 0.1349\\
\hline $2_3^+$ & 0 & 0.1349 & 0\\

\end{tabular}
\caption{Mixing matrix elements for the first three unperturbed $0^+$ and $2^+$ states. The matrix elements are expressed in MeV.}\label{tab:mmelements}
\end{center}
\end{table}
We now make the approximation to neglect the mixing between the lowest and first excited $0^+$ states. We can make this approximation because the mixing matrix element coupling those two states is less than half the magnitude of the interaction matrix element coupling the $0^+_2$ and $0^+_3$ states. This is strengthened by the fact that the energy difference between the $0^+_1$ and $0^+_2$ states is much larger than between the $0^+_2$ and $0^+_3$ states. The situation thus reduces to a simple two level mixing problem for which the energies and eigenvectors are easily determined \cite{KHEY1990}. We obtain a level at 494 keV and one at 882 keV, compared to the absolute energies 489 keV and 853 keV resulting from the full diagonalisation. Although we neglected the effect of the other 55 $0^+$ levels, we obtain a good result for the energies using two-level mixing only. This is also reflected in the wavefunctions. According to the two-level mixing problem, the wavefunctions look like
\begin{eqnarray}
\psi_1&=&a_{11}\phi^{(0)}_1 + a_{21}\phi^{(0)}_2~,\\
\psi_2&=&a_{12}\phi^{(0)}_1 + a_{22}\phi^{(0)}_2~,
\end{eqnarray}
with
\begin{equation}
a_{11}=\mp a_{22}~,
a_{12}=\pm a_{21}~.\label{eq:magnitude}
\end{equation}

\noindent Table \ref{tab:percentages} gives the coefficients for the wavefunctions of the $0^+$ (and $2^+$) states resulting from the full diagonalisation expressed in the unperturbed basis \eqref{eq:intermediate}. Only the coefficients of the lowest three unperturbed $0^+$ (and $2^+$) states are shown. Inspecting the coefficients of the unperturbed 2p-2h and 4p-4h $0^+$ wavefunctions $|0_1^+\rangle'_{N+2}$ and $|0_1^+\rangle'_{N+4}$ respectively for the $0_2^+$ and $0_3^+$ wavefunctions, we find the same structure for the magnitudes as in equation \eqref{eq:magnitude}. One notices that a simple two-level mixing approximation cannot reproduce the phases resulting from the full diagonalisation. Calculation of the coefficients in the two-level approximation for the $0^+$ states results in the values 0.7696 and 0.6384, which are in good agreement with the magnitudes resulting from the full diagonalisation.
\begin{table}[!htb]
\begin{tabular}{l|l|l|l}
\hline\hline ~ & $|0_1^+\rangle'_N$ & $|0_1^+\rangle'_{N+2}$ & $|0_1^+\rangle'_{N+4}$ \\
\hline\hline $|0_1^+\rangle$ & -0.9897 & -0.1310 & 0.0309\\
\hline $|0_2^+\rangle$ & 0.1216 & -0.7624 & 0.6148\\
\hline $|0_3^+\rangle$ & -0.0611 & 0.5893 & 0.7655\\
\hline\hline
\end{tabular}
\begin{tabular}{l|l|l|l}
\hline\hline ~ & $|2_1^+\rangle'_N$ & $|2_1^+\rangle'_{N+2}$ & $|2_1^+\rangle'_{N+4}$ \\
\hline\hline $|2_1^+\rangle$ & 0.1907 & -0.6379 & 0.7091 \\
\hline $|2_2^+\rangle$ & 0.5964 & -0.4906 & -0.5664 \\
\hline $|2_3^+\rangle$ & 0.7696 & 0.5114 & 0.2187 \\
\hline\hline
\end{tabular}
\caption {Coefficients for the wavefunctions of the $0^+$ (and $2^+$) states resulting from the full diagonalisation expressed in the unperturbed basis \eqref{eq:intermediate}. Only the coefficients of the lowest three unperturbed states $|0^+_1\rangle'_\nu$ and $|2^+_1\rangle'_\nu$ ($\nu=N,N+2,N+4$) are given.}\label{tab:percentages}
\end{table}
\\\\\noindent  The above results indicate that the two-level approximation to the full diagonalisation is a very good one. It thus follows that a knowledge of the mixing matrix elements and the energies of the unperturbed levels can lead to a better understanding in the process of configuration mixing. We like to point out that this severe approximation cannot reproduce the correct phase, so we cannot use these approximated wavefunctions for the calculations of B(E2)-values.
\section{B(E2)-values and construction of collective bands}\label{sect:be2values}
\noindent In the process of separate diagonalisation of the full Hamiltonian \eqref{eq:hamiltonian} (excluding $\hat{V}_{mix}$) in the three subspaces separately, we have also made an 'intermediate' calculation concerning B(E2)-values. The knowledge of the basis states \eqref{eq:full} and \eqref{eq:intermediate} gives rise to two interesting expressions for the reduced matrix element of a transition between the initial $J_i(i)$ and final $J_f(f)$ state, i.e.
\begin{align}
&\langle J_f,f||T(E2)||J_i,i\rangle=\nonumber\\
&\sum_k\sum_p a^\nu_{k,i}(J_i)a^\nu_{p,f}(J_f)~_{\nu}\langle J_f,p||T(E2)||J_i,k\rangle_\nu\Big|_{\nu=N}\nonumber\\
&+\sum_l\sum_q a^{\nu}_{l,i}(J_i)a^{\nu}_{q,f}(J_f)~_{\nu}\langle J_f,q||T(E2)||J_i,l\rangle_{\nu}\Big|_{\nu=N+2}\nonumber\\
&+\sum_m\sum_r a^{\nu}_{m,i}(J_i)a^{\nu}_{r,f}(J_f)~_{\nu}\langle J_f,r||T(E2)||J_i,m\rangle_{\nu}\Big|_{\nu=N+4}~,\label{eq:nr1}
\end{align}
and,
\begin{widetext}
\begin{align}
\langle J_f,f||T(E2)||J_i,i\rangle=&\sum_{k,p,s,s'}a^\nu_{k,i}(J_i)a^\nu_{p,f}(J_f)\tilde{b}^{\nu}_{s,k}(J_i)\tilde{b}^{\nu}_{s',p}(J_f)~_\nu'\langle J_f,s'||T(E2)||J_i,s\rangle'_\nu\Big|_{\nu=N}\nonumber\\
&+\sum_{l,q,t,t'}a^{\nu}_{l,i}(J_i)a^{\nu}_{q,f}(J_f)\tilde{b}^{\nu}_{t,l}(J_i)\tilde{b}^{\nu}_{t',q}(J_f)~'_{\nu}\langle J_f,t'||T(E2)||J_i,t\rangle'_{\nu}\Big|_{\nu=N+2}\nonumber\\
&+\sum_{m,r,u,u'}a^{\nu}_{m,i}(J_i)a^{\nu}_{r,f}(J_f)\tilde{b}^{\nu}_{u,m}(J_i)\tilde{b}^{\nu}_{u',r}(J_f)~'_{\nu}\langle J_f,u'||T(E2)||J_i,u\rangle'_{\nu}\Big|_{\nu=N+4}~,\label{eq:nr2}
\end{align}
\end{widetext}
with $\tilde{b}_{s,k}$ component of transposed matrix $\mathbf{\tilde{B}}$ of equation \eqref{eq:intermediate}.
Expression \eqref{eq:nr2} is most interesting because it allows us to check which transitions in the unperturbed bands make up for an important contribution to a certain transition $J_i(i)\rightarrow J_f(f)$. We now apply the above method to the particular case of $^{188}$Pb.\\\\
For $^{188}$Pb, only two experimental B(E2)-values are known \cite{AWAL2003}
\begin{eqnarray}
&&B(E2;2_1^+\rightarrow 0_1^+)=5(3) \textrm{~W.u.}~,\\
&&B(E2;4_1^+\rightarrow 2_1^+)=160(80) \textrm{~W.u.}~.
\end{eqnarray}
In the calculation of the E2-transition rates, we use the consistent-Q procedure \cite{DWAR1983} to determine the E2-transition operator as
\begin{equation}
T(E2)=\sum_{i=1}^{3} e_i \left[\left(s^{\dag}\tilde{d}+d^{\dag}\tilde{s}\right)^{(2)}+\chi_i\left(d^{\dag}\tilde{d}\right)^{(2)}\right]~.
\end{equation}
So we choose the values for $\chi_{2p-2h}$ and $\chi_{4p-4h}$ as obtained in \cite{RFOS2003} and fit the effective charge to those two known data. We took $e_{2p-2h}$ and $e_{4p-4h}$ 1.2 times $e_{reg}$. Table \ref{tab:e2} lists the parameters used. We point out that we have chosen $\chi_{reg}=0$. We made this choice because there are no data available for transitions within the regular band, hence we are not able to fit $\chi_{reg}$ to known experimental values.  Evidently, the groundstate is regular and we know the B(E2)-value for the $2_1^+\rightarrow 0^+_1$ transition, but as $\chi_{reg}$ is of no influence to $2^+\rightarrow 0^+$ transitions in the $U(5)$-limit, we cannot make use of this transition in determining $\chi_{reg}$.
\begin{table}[!tb]
\begin{center}
\begin{tabular}{ccccc}
\hline\hline $\chi_{reg}$& $\chi_{2p-2h}$ & $\chi_{4p-4h}$ & $e_{reg}$ & $e_{2p-2h}=e_{4p-4h}$\\
\hline 0 & 0.515& -0.680 & 0.11 & 0.132\\
\hline\hline
\end{tabular}
\caption{Parameters for E2-transitions in $^{188}$Pb. The effective charges are expressed in $e\cdot b$, the $\chi$ are dimensionless.}\label{tab:e2}
\end{center}
\end{table}
\\\\For this choice of parameters the IBM calculation yields the following results
\begin{eqnarray}
&&B(E2;2_1^+\rightarrow g.s.)=0.0195~ e^2b^2~ \textrm{or}~ 3~ \textrm{W.u.}~,\\
&&B(E2;4_1^+\rightarrow 2_1^+)= 0.9747~ e^2b^2~ \textrm{or}~ 152~ \textrm{W.u.}~.
\end{eqnarray}
\begin{figure}[!htb]
\includegraphics [width=8.6cm]{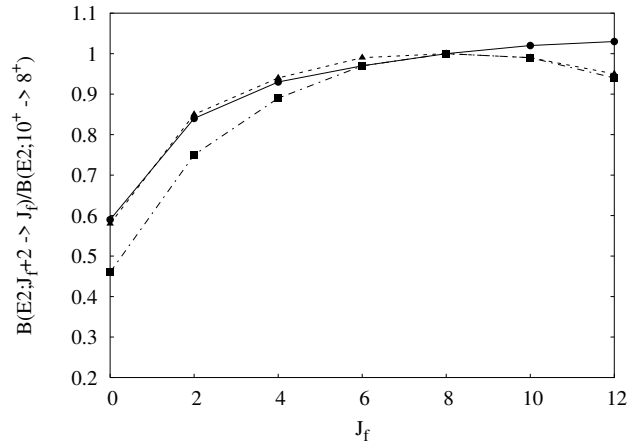}
\caption{The ratio $B(E2;J_f+2\rightarrow J_f)/B(E2;10^+\rightarrow 8^+)$ in the unperturbed 2p-2h and 4p-4h bands compared to the theoretical estimate for a pure rotational band. The full line denotes the theoretical estimate, the dot dashed line the ratio for the 2p-2h unperturbed band and the dotted line the ratio in the 4p-4h unperturbed band.}\label{fig:gsband}
\end{figure}

\noindent The fact that the parameters are able to reproduce two B(E2)-values that differ two orders of magnitude, permits us to make further theoretical predictions.\\\\
To begin with, it is interesting to make a comparison between the B(E2)-values for the unperturbed 2p-2h and 4p-4h band and the theoretical estimate using a pure rotational band \cite{BOMO1975,WGRE1996}
\begin{align}
B(E2&;J_f+2~\textrm{g.s.band}\rightarrow J_f~\textrm{g.s.band})\nonumber\\&= C \left(2I_f+1\right)\left(\begin{array}{ccc}J_f+2 &       2&J_f\\0&0&0\end{array}\right)^2 ~,\\
&=C \frac{3}{2}\frac{(J_f+1)(J_f+2)}{(2J_f+3)(2J_f+5)}~,\label{eq:nr3}
\end{align}
with C a constant factor depending on the nucleus. Since C is unknown, we can only compare with the B(E2)-values relative to a certain transition. In Figure \ref{fig:gsband} we illustrate the $J_f$-dependent part of equation \eqref{eq:nr3} and draw a parallel with the B(E2)-values in the unperturbed bands, all relative to the $10^+\rightarrow 8^+$ E2-transition.

One can clearly see that the band built upon the unperturbed 4p-4h $0^+$ state follows the rotational structure very well, while the band constructed upon the unperturbed 2p-2h $0^+$ exhibits larger deviations, especially for the transitions between the lower spin states. Because $\chi_{4p-4h}$ was fixed using I-spin symmetry \cite{RFOS2003}, we can explain the rotational-like behaviour of the unperturbed 4p-4h band by comparing with the regular band in $^{180}$W. This band is recognised as a $K^\pi=0^+$ rotational groundstate band \cite{NDAT1994}, which explains the good agreement of the slope of the ratio $B(E2;J_f+2\rightarrow J_f)/B(E2;10^+\rightarrow 8^+)$ in the unperturbed 4p-4h band with the theoretical one.\\\\
\begin{table}
\begin{tabular}{c|c}
\hline\hline transition & B(E2)-value\\
\hline\hline $4_2^+\rightarrow 2_3^+$ & 44 W.u.\\
\hline $2_3^+\rightarrow 2_1^+$ & 36 W.u.\\
\hline $2_3^+\rightarrow 2_2^+$ & 2 W.u.\\
\hline\hline
\end{tabular}\caption{Inter-band transitions involving the $2_3^+$ state}\label{tab:twoplus}
\end{table}
\begin{figure*}[!htb]
\begin{center}
\includegraphics [width=10cm,angle=-90]{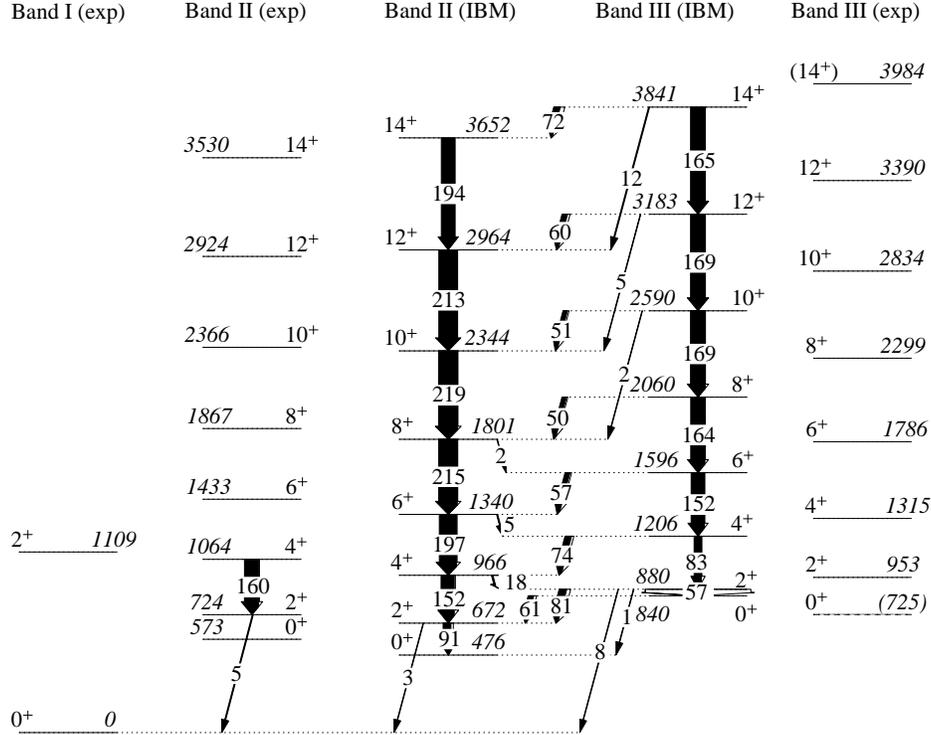}
\end{center}
\caption{Experimental and theoretical level scheme in $^{188}$Pb. The arrows denote the B(E2)-values for a given transition, expressed in W.u. The experimental data were taken from \cite{GDRA2004,AWAL2003,GDRA2003,YLEC1999}.}\label{fig:levelscheme}
\end{figure*}
\noindent Finally, we come to the question about labelling the mixed states, resulting from diagonalisation of the full Hamiltonian \eqref{eq:hamiltonian}, in a meaningful way into a given band. As the criterium, we start from the calculated high-spin members of the two bands and follow the E2-decay. We define the collective band through a sequence of E2-transitions down to low spin such that the intra-band B(E2)-values are bigger than the inter-band B(E2)-values. Moreover in the case of $^{188}$Pb, we have as a second criterium the fact that the intra-band transitions between higher spin (and thus relatively unperturbed) states should exhibit an approximately constant behaviour as illustrated in Fig. \ref{fig:gsband}. The final results are shown in Figure \ref{fig:levelscheme}. In Table \ref{tab:twoplus} all inter-band transitions involving the IBM $2_3^+$ at 969 keV (not drawn in Fig. \ref{fig:levelscheme}) are shown. The transitions between Band II and Band III in Figure \ref{fig:levelscheme} that are not shown, are less than 1 W.u. We like to stress that the 725 keV $0^+$ state in Figure \ref{fig:levelscheme} is indicated with a dashed line and tentative energy. This state was identified in two experimental studies \cite{RALL1998,YLEC1999}, but recently performed experiments \cite{GDRA2004,KVEL2003} do not reach conclusive evidence for this state and its energy. The position and existence of this level thus is still under discussion.
\\\\
Comparing the IBM results with the experimental bands, we notice that the structure of the collective bands is reproduced rather well. The energy levels that constitute the more collective Band II are in good agreement with the experimental results. The structure of the Band III is also described rather well although the value of $\Delta_{2p-2h}$ seems a bit too small and the mixing parameter between the two intruder configurations slightly too large. Although the $0_2^+$ and the $0_3^+$ state are placed in the collective bands the other way round in reference \cite{AWAL2003}, here we switched them. This change was performed following the definition that inter-band transitions must be small compared to intra-band transitions. In the other case than the one  depicted here, the IBM-calculations give transitions from the collective Band III to the regular ground state that are larger and thus more collective than the transition to the $0^+$ state of Band III itself. Such a decay pattern should definitely be discarded. The only way to obtain a consistent picture in which the E2-transitions exhibit a collective nature within the collective bands and give rise to small E2-transitions between the different bands, is by associating the $0^+_2$ level to Band II and the $0_3^+$ level to Band III.\\\\
When we take a closer look at the E2-transitions between the low-lying levels, something interesting occurs. Using expression (\ref{eq:nr2}), one can single out those contributions of the reduced matrix elements for the transitions in the unperturbed bands (thus reduced matrix element and coefficient!), that make up the major contribution to a given transition $J_i(i)\rightarrow J_f(f)$ with the condition that their sum may not deviate more than 10\% from the total value with all contributions taken into account. 
\begin{figure}[!htb]
\includegraphics [height=8.1cm,angle=-90] {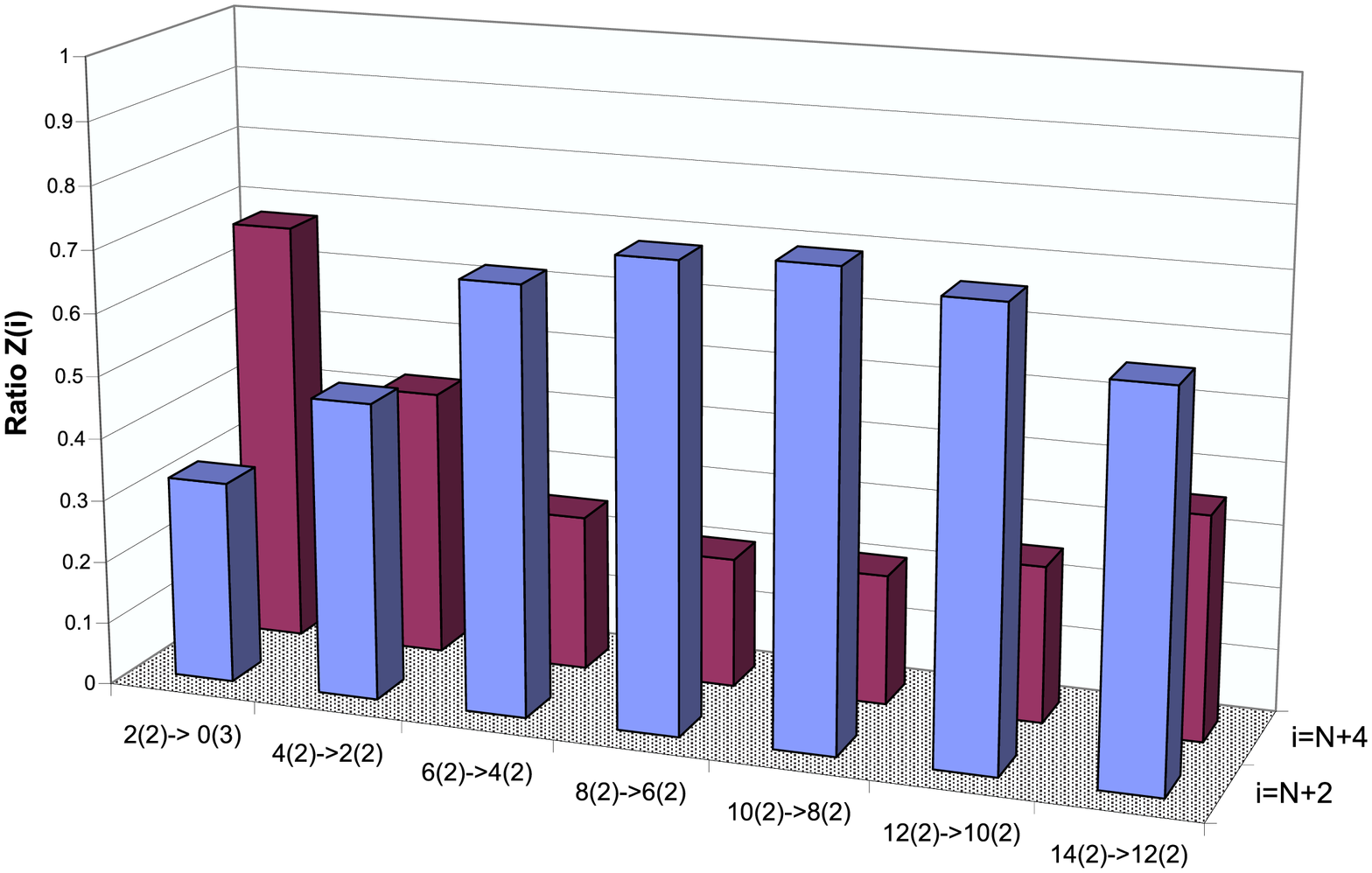}
\includegraphics [height=8.1cm,angle=-90]{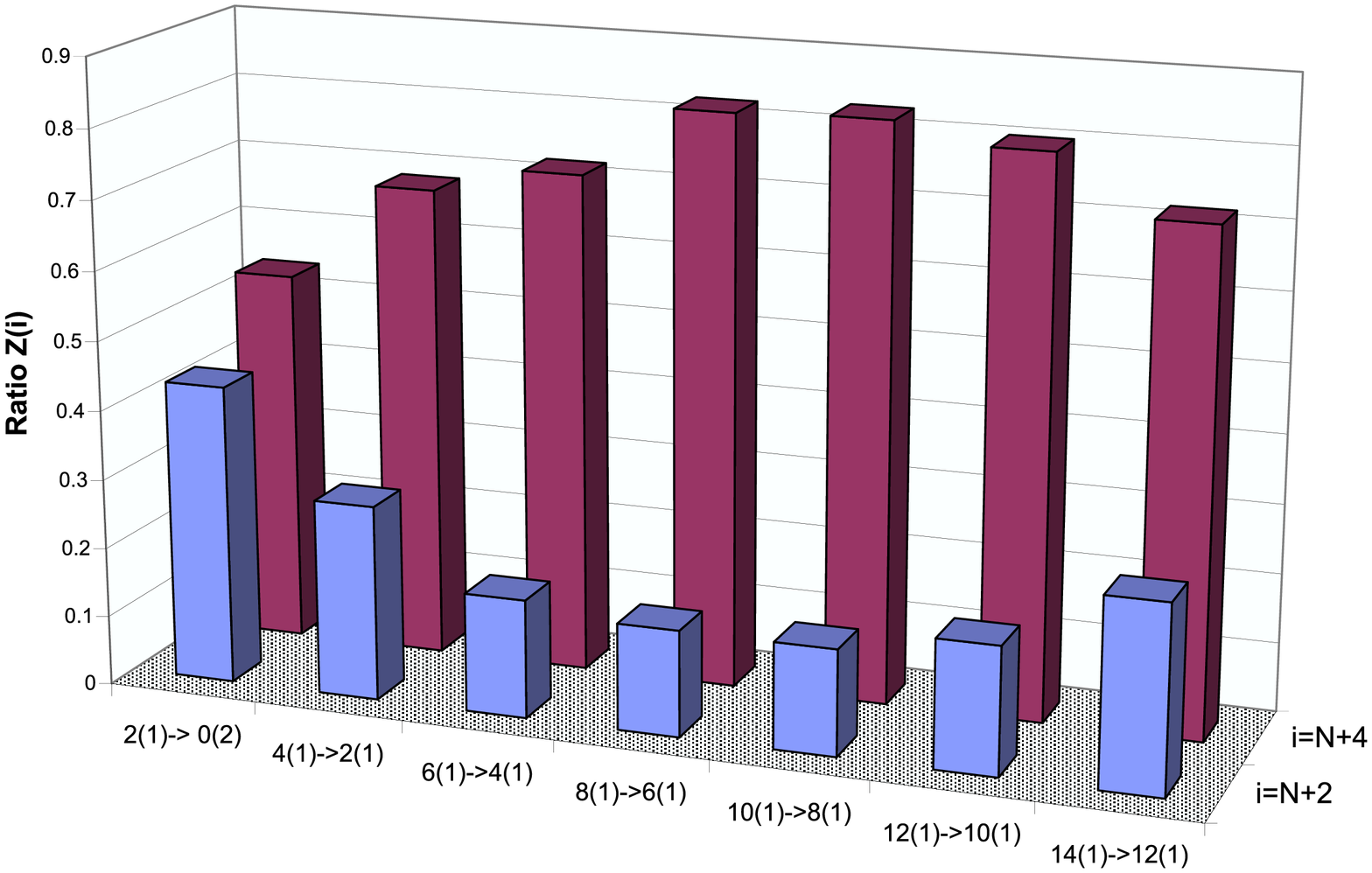}
\caption{(colour online) The left part shows the ratio's $Z(N+2)$ and $Z(N+4)$ (see equation \eqref{eq:percentage}) for the  intra-band transitions in Band III, the right part shows the ratio $Z(N+2)$ and $Z(N+4)$ for the intra-band transition in Band II. }\label{fig:contributions}
\end{figure}
In the case of $^{188}$Pb, it turns out that for the so constructed Band II and Band III, the main contribution to intra-band transitions always consists of the term with the corresponding lowest rank transition in the unperturbed 2p-2h band and the one in the unperturbed 4p-4h band. Thus
\begin{widetext}
\begin{align}
&\langle J_f,f||T(E2)||J_i,i\rangle \cong\nonumber\\
&\sum_{l,q} a_{l,i}^{N+2}(J_i)a^{N+2}_{q,f}(J_f)\tilde{b}_{1,l}^{N+2}(J_i)\tilde{b}^{N+2}_{1,q}(J_f)~_{N+2}'\langle J_f,1||T(E2)||J_i,1\rangle'_{N+2} \nonumber\\
&+\sum_{m,r} a_{m,i}^{N+4}(J_i)a^{N+4}_{r,f}(J_f)\tilde{b}_{1,m}^{N+4}(J_i)\tilde{b}^{N+4}_{1,r}(J_f)~_{N+4}'\langle J_f,1||T(E2)||J_i,1\rangle'_{N+4}\nonumber\\
&= Z(N+2)\langle J_f,f||T(E2)||J_i,i\rangle + Z(N+4)  \langle J_f,f||T(E2)||J_i,i\rangle
\label{eq:percentage}
\end{align}
\end{widetext}
with $|J_i,i\rangle$ and $|J_f,f\rangle$ both states in Band II or both states in Band III. This fact is as expected. More interesting is the ratio $Z$ of those two main contributions to the total value of the reduced matrix element. These results are depicted in Figure \ref{fig:contributions}.\\\\
Notice that the ratio $Z(N+4)$ becomes larger than the ratio $Z(N+2)$ for the $2_2^+\rightarrow0_3^+$ transition in Band III. This seems to be in contradiction with the fact that we placed the  $2_2^+$ and  $0_3^+$ states in Band III as for all other transitions in this bands $Z(N+2)$ dominates over $Z(N+4)$. However, all other orderings of these levels into the bands (II and III) are in contradiction with the E2-decay pattern and our definition of intra-band and inter-band transitions strengths in order to define a collective band. 
This very large ratio $Z(N+4)$ for the $2_2^+\rightarrow0_3^+$ transition in Band III is due to the combined effect of the large mixing and the relative magnitudes of the two contributing reduced matrix elements \eqref{eq:percentage}.
\\\\
So we can conclude that the only physically meaningful way to characterise the low-lying states is by constructing the different bands using the intensities of the sequences of E2-transitions. Attempts to label the extremely mixed low-lying states as '2p-2h' or '4p-4h' lead to inconsistencies.
\section{Comparison with the collective rotational model}\label{sect:rotor}
\noindent Having constructed two different collective bands starting from the IBM and using the prescription to define bands on the basis of the calculated B(E2)-values, it is very useful to check for consistency with the results of other theoretical approaches. First of all we concentrate on the quadrupole moments.\\\\
Within the IBM, the quadrupole moments are calculated as
\begin{align}
Q(J)&= \sqrt{\frac{16\pi}{5}}\langle JJ|T(E2)_0|JJ\rangle\nonumber\\
&=\sqrt{\frac{16\pi}{5}}\left(\begin{tabular}{ccc} $J$ & $2$ & $J$ \\ $-J$ & $0$ & $J$ \end{tabular}\right ) \langle J || T(E2)|| J \rangle~.
\end{align}
From the point of view of the collective rotational model\cite{BOMO1975,WGRE1996,DROW1970}, the electric quadrupole moment is defined as 
\begin{align}
Q=\sqrt{\frac{16\pi}{5}}\langle J,K,M=J|\mathcal{M}(E2,0)|J,K,M=J\rangle\label{eq:quadrupole}~.
\end{align}
For $K=0$ bands this reduces to
\begin{align}
Q=\frac{-J(J+1)}{(J+1)(2J+3)}Q_0^0~,\label{eq:intquadrupole}
\end{align}
with the intrinsic quadrupole moment $Q_0^0$ defined as
\begin{align}
Q_0^0= \sqrt{\frac{16\pi}{5}}\langle K=0|\mathcal{M}(E2,0)|K=0\rangle~.
\end{align}
Equating the quadrupole moments of equation \eqref{eq:quadrupole} and \eqref{eq:intquadrupole} allows us to extract an equivalent intrinsic quadrupole moment $Q_0^0$. Likewise one can use the IBM B(E2)-values to extract an equivalent intrinsic quadrupole moment $Q_0^0$ using the collective rotational model B(E2)-expression
\begin{align}
B(E2;(J+2),K=0&\rightarrow J,K=0)=\nonumber\\
&\frac{5}{16\pi} (2J+1)\left(\begin{tabular}{ccc}$J+2$& $2$ & $J$\\ $0$&$0$&$0$\end{tabular}\right)^2 (Q_0^0)^2~.
\end{align}
Having extracted an intrinsic quadrupole moment, one can deduce a deformation parameter $\beta_0$ using the expression
\begin{align}
Q_0=\frac{3ZR_0^2}{\sqrt{5\pi}}\beta_0~.
\end{align}
If the concept of a collective band characterised by a single $Q_0^0$ value is to make sense, one expects that values of $Q_0^0$ extracted with these two procedures will not differ much. Moreover, one expects only a moderate variation of $Q_0^0$ as a function of J along the band.

\begin{figure}[!htb]
\includegraphics [width=8.6cm] {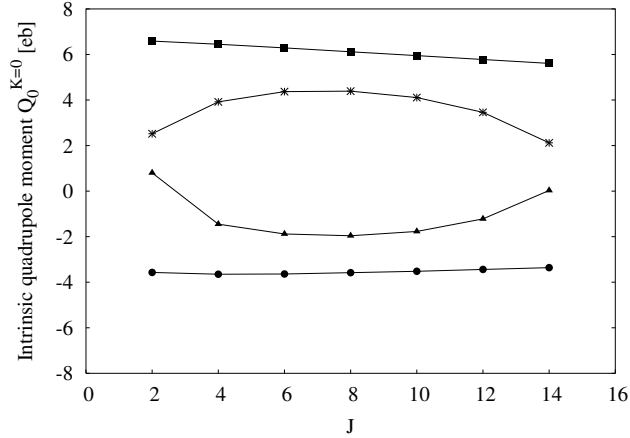}
\caption{Intrinsic quadrupole moment extracted from transition matrix elements, calculated within the IBM. The intrinsic quadrupole moment is expressed in $e\cdot b$. Results for the 2p-2h unperturbed band are represented with $\bullet$, results for the 4p-4h unperturbed band with $\blacksquare$. Bands II and III are represented respectively by $\ast$ and $\blacktriangle$.}\label{fig:iqm1}
\end{figure}

In Figure \ref{fig:iqm1}, we have plotted the intrinsic quadrupole moment $Q_0^0$ derived from the E2 diagonal matrix elements. We clearly see that the intrinsic quadrupole moment stays approximately constant for the unperturbed bands. The intrinsic quadrupole moments of the unperturbed 4p-4h band are positive, indicating a prolate deformation of the nucleus, while the negative sign of the intrinsic quadrupole moments in the unperturbed 2p-2h band is consistent with an oblate deformation. The values of the intrinsic quadrupole moments for the unperturbed bands obtained from the IBM diagonal transition matrix elements are in good agreement with the quadrupole moment magnitudes reported by Dracoulis \textit{et al.} \cite{GDRA2004}. They calculated a magnitude of 3.4 eb for the intrinsic quadrupole moment of the unperturbed oblate band, while the IBM gives an intrinsic quadrupole moment varying between -3.4 eb and -3.7 eb. The same is valid for the unperturbed prolate band with a quadrupole moment of 6.2 eb resulting from their band mixing and a quadrupole moment varying between 5.6 eb and 6.6 eb resulting from the IBM-calculations. The mixing causes the bands to become less prolate and oblate so the intrinsic quadrupole moments  for Band II and III respectively are considerably smaller, as can be seen in Figure \ref{fig:iqm1}.\\\\
\begin{figure}[!h]
\includegraphics [width=8.1cm] {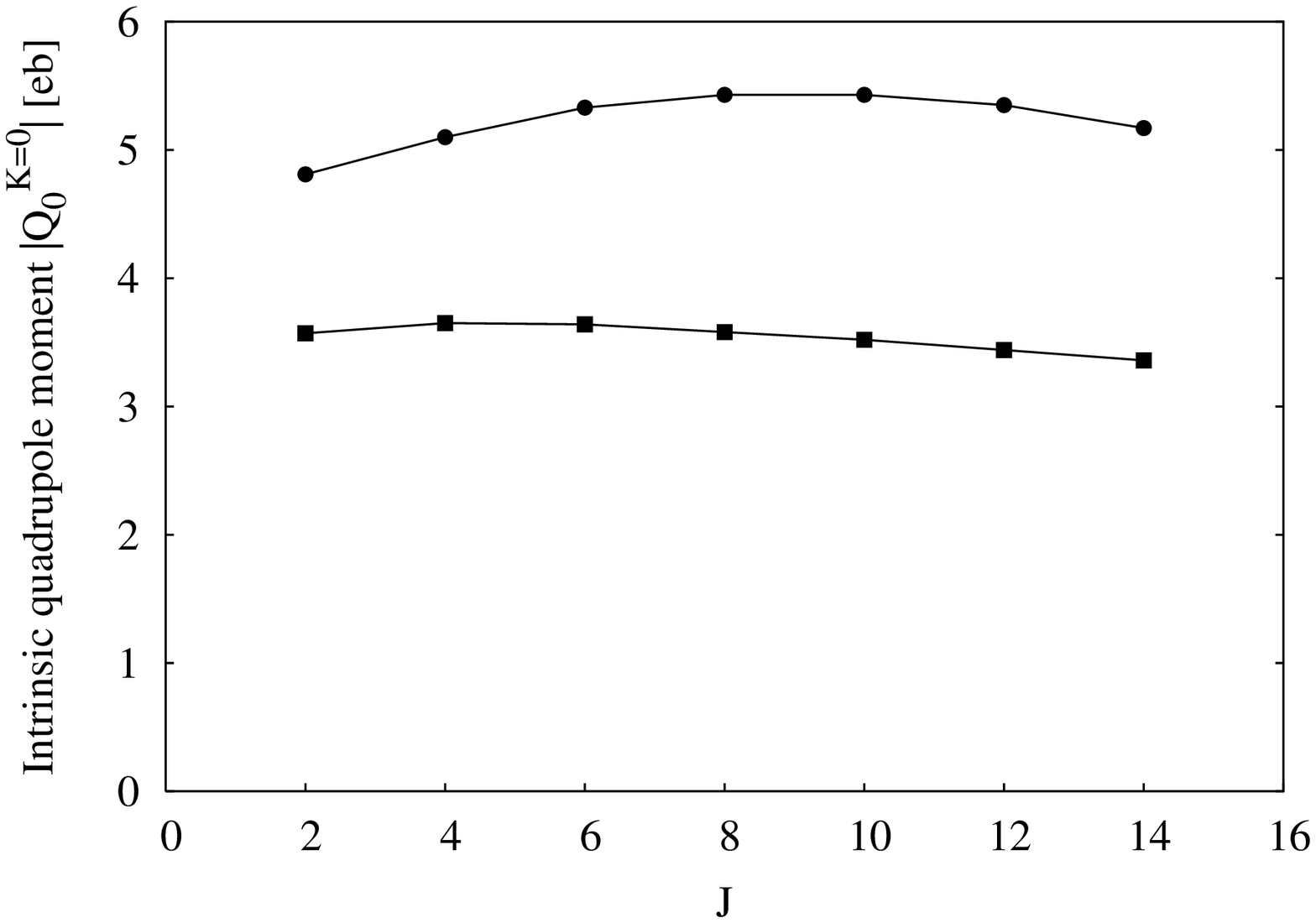}
\includegraphics [width=8.1cm]{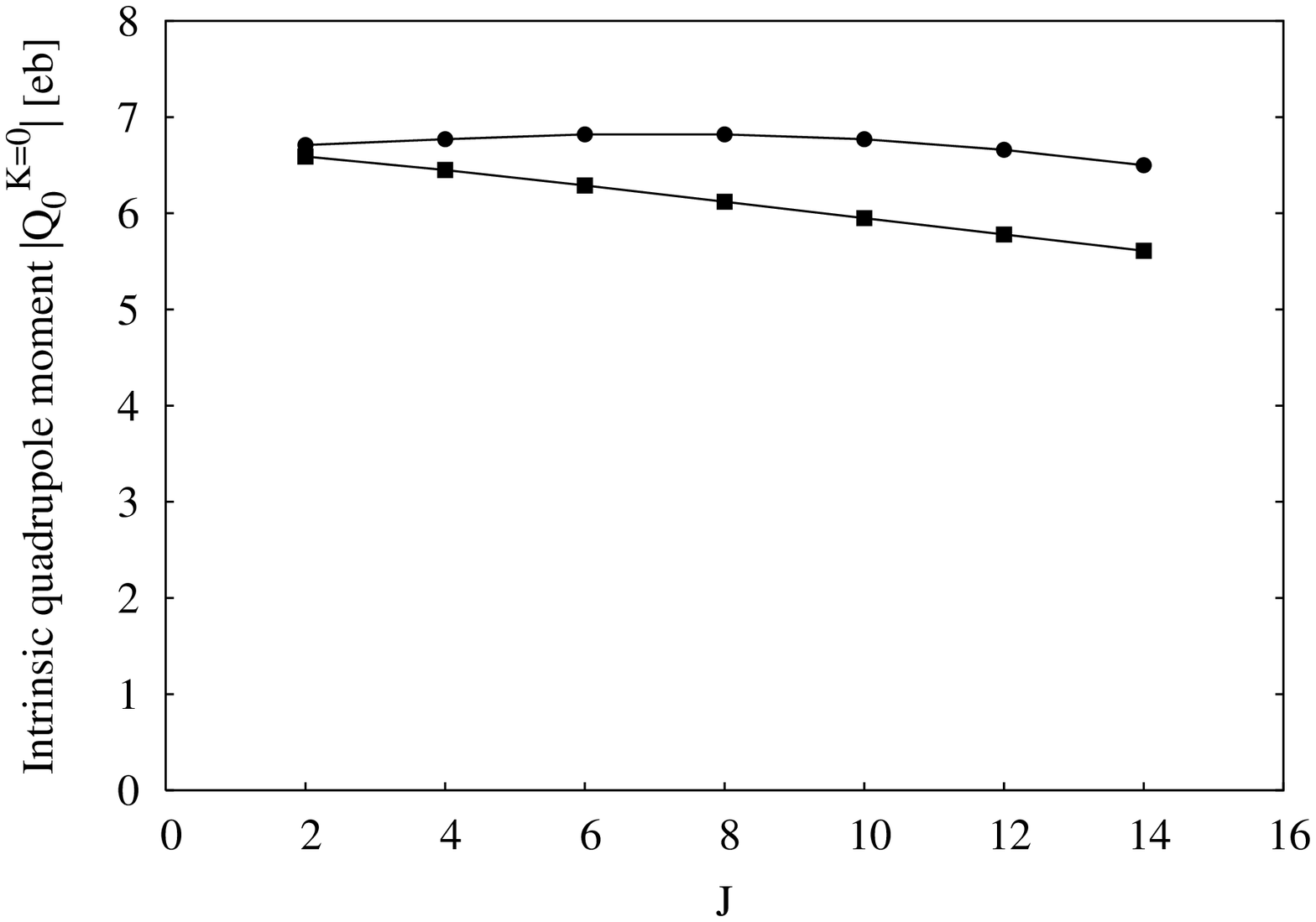}
\caption{The upper figure shows the comparison between the magnitudes of the intrinsic quadrupole moments extracted from the theoretical B(E2)-values ($\bullet$) and the magnitudes of the intrinsic quadrupole moments extracted from the diagonal matrix elements ($\blacksquare$), both for the unperturbed 2p-2h band. The lower figure shows the same comparison for the unperturbed 4p-4h band.}\label{fig:iqm2}
\end{figure}
Extracting intrinsic quadrupole moments from B(E2)-values gives a similar picture as in Fig.~\ref{fig:iqm1}. More interesting is the comparison between the intrinsic quadrupole moments calculated from the B(E2)-values and the absolute values of the intrinsic quadrupole moments as extracted from equations \eqref{eq:quadrupole} and \eqref{eq:intquadrupole}. Figure \ref{fig:iqm2} depicts the comparison for the unperturbed 2p-2h (upper part) and the unperturbed 4p-4h band (lower part). It can be clearly seen that the unperturbed 4p-4h band exhibits a rotational-like behaviour to large extent, while the unperturbed 2p-2h band is less rotational. This is in agreement with our conclusion for Fig.~\ref{fig:gsband}.
\\
Finally, we have also extracted the deformation parameter $\beta_0$. For the unperturbed 2p-2h band $\beta_0$ varies between -0.11 and -0.12, while for the unperturbed 4p-4h band $\beta_0$ stays between 0.19 and 0.22. These values are smaller than the deformation parameters calculated by Bender \textit{et al.} \cite{MBEN2004}.
\section{Conclusion}
\noindent In the present paper we have carried out a detailed analysis of the presence of various families of collective bands, in particular for $^{188}$Pb. We have started from an algebraic model approach (the interacting boson model) in order to truncate the extended shell-model space that incorporates the presence of proton particle-hole excitations across the $Z$=82 closed shell. Moreover, making use of a concept called intruder symmetry, we have been able to reduce the number of parameters in the present description. A first calculation has been carried out by Fossion \textit{et al.}\cite{RFOS2003}, accentuating the presence of three different families. Here, we have defined an intermediate basis that defines three separate systems by diagonalizing the Hamiltonian in the 0p-0h, 2p-2h, and 4p-4h configuration spaces separately. This basis allows to understand the mixing between the unperturbed bands in a transparent way. Moreover, we have used the E2-decay, starting at high spin, to define "physical" bands also progressing to low-spin members. Here the conclusion points towards an important mixing between the $0_2^+$, $0_3^+$ and the $2_1^+$, $2_2^+$ and $2_3^+$ band members, still allowing the separation into two collective band structures. A simple reanalysis of these two bands within the collective rotational model is consistent with  prolate and oblate band characteristics for the unperturbed 4p-4h and 2p-2h bands respectively. Extracted magnitudes of the intrinsic quadrupole and the deformation parameters are in good agreement with calculations starting from mean-field approaches.\\\\
It is our aim to carry out a similar analysis for the other Pb-nuclei near the neutron mid-shell region and follow the mixing patterns between the two bands when moving away from the mid-shell region (at N=104) and to incorporate an extensive study of electromagnetic properties (E2- and E0-transitions) and the study of isotopic and isomeric shifts as well. 
\section{ACKNOWLEDGEMENTS}
\noindent The authors are most grateful to G.~Dracoulis and R.~Julin for extensive discussions and to R.~Janssens, R.~Page, A.~Andreyev, A.~Dewald  for comments on the present manuscript. They also like to thank P.-H.~Heenen, M.~Bender, P.~Van Isacker, P.~Van Duppen, and J.L.~Wood for their continuous interest in this research. Financial support from the ``FWO-Vlaanderen'' (V.H. and K.H.), the University of Ghent (S.D.B. and K.H.) and the IWT(R.F.) as well as from the OSTC (Grant IUAP \#P5/07) is acknowledged.

\end{document}